\documentclass{ws-ijmpa}
\usepackage[super,compress]{cite}
\usepackage[colorlinks=true, pdfstartview=FitV, linkcolor=red, citecolor=blue, urlcolor=blue]{hyperref}
\usepackage{graphicx}
\usepackage{booktabs}
\begin{document}
\markboth{A. Monnai}{Direct photons in hydrodynamic modeling of relativistic nuclear collisions}

%
\catchline{}{}{}{}{}
%

\title{DIRECT PHOTONS IN HYDRODYNAMIC MODELING OF RELATIVISTIC NUCLEAR COLLISIONS
}

\author{AKIHIKO MONNAI}

\address{
Department of Mathematical and Physical Sciences, Japan Women's University\\ 
Bunkyo-ku, Tokyo 112-8681, Japan
\\
monnaia@fc.jwu.ac.jp}

\maketitle

\begin{history}
\received{\today}
\revised{Day Month Year}
\end{history}

\begin{abstract}
We review direct photons in the phenomenology of nuclear collisions at relativistic energies. Direct photons carry information about the space-time evolution of nuclear collisions because the QCD medium is transparent against colorless particles. After a status summary of theoretical and experimental studies, transverse momentum spectra and azimuthal momentum anisotropies of direct photons are studied in the context of the hydrodynamic modeling of relativistic nuclear collisions with emphasis on pre-equilibrium photons. It is implied that pre-equilibrium photons can be as important as thermal and prompt photons for comprehensive understanding of direct photon production.

\end{abstract}

\ccode{PACS numbers: }


\section{Introduction}	

Quarks and gluons are elementary particles described by quantum chromodynamics (QCD), the theory of the strong interaction. The many-body system of quarks and gluons is known to have various phases under different temperatures and chemical potentials. One of the first phase diagrams of QCD was conjectured back in 1970's by Cabbibo and Parisi \cite{Cabibbo:1975ig} and have been kept updated \cite{Baym:1982sn,Baym:1984hpn,Hatsuda:1985eb,Asakawa:1989bq,Alford:1998mk,McLerran:2007qj,Fukushima:2008wg}. The system is normally confined to hadrons, composite particles of quarks and gluons such as protons and neutrons found in nuclei. On the other hand, it is expected to be deconfined to the quark-gluon plasma (QGP) at high temperatures and densities \cite{Letessier:2002gp,Rafelski:2003zz,Yagi:2005yb,Wang:2016opj}. Numerical results of (2+1)-flavor lattice QCD simulations indicate that the quark-hadron transition is a crossover \cite{Aoki:2006we} and it occurs around $T_c \sim 155$-$160$ MeV in the limit of vanishing chemical potentials \cite{Aoki:2009sc, Bazavov:2011nk, Borsanyi:2013bia, Bazavov:2014pvz}.

Relativistic heavy-ion colliders have been a powerful tool for experimentally understanding the properties of the QCD matter. LBNL Bevalac is one of the earliest facilities to run such collisions, followed by BNL Alternating Gradient Synchrotron (AGS) and CERN Super Proton Synchrotron (SPS). The first evidence of the deconfined quark matter was found at BNL Relativistic Heavy Ion Collider (RHIC) \cite{Adcox:2004mh,Adams:2005dq,Back:2004je,Arsene:2004fa}. This has been a major advancement in the field of QCD for several reasons. First, it is the first (pseudo-)phase transition of a quantum field of elementary particles that we have observed in laboratories, which takes us closer to the state of matter at the beginning of the Universe. Second, quantitative comparisons of theoretical estimations and experimental data can be performed, opening the door to the QCD matter physics as precision physics. The heavy-ion programs at CERN Large Hadron Collider (LHC) have later confirmed that the QGP is produced at much higher energies \cite{Aamodt:2010pa,ATLAS:2011ah,Chatrchyan:2012wg}, and the beam-energy scan programs at BNL RHIC and CERN SPS now probe threshold collision energy for the QGP production along with the phase structure of QCD at finite chemical potentials.

An essential discovery in QCD collider experiments is that hadronic particle yields and their momentum dependences are well described by the relativistic hydrodynamic model \cite{Kolb:2000fha,Schenke:2010rr}. Fourier coefficients of the observed transverse momentum spectra, flow harmonics $v_n$, strongly reflect the spatial anisotropy of the collision geometry, and are consistent with the results of hydrodynamic simulations. Elliptic flow $v_2$ was the first to be investigated because of the oval-shaped overlapping region of two nuclei \cite{Ollitrault:1992bk, Poskanzer:1998yz}, and higher order harmonics, including triangular flow $v_3$, started to be a topic of interest as event-by-event fluctuations attracted attention \cite{Takahashi:2009na,Alver:2010gr}. Those are strong evidences that the system is strongly-coupled and behaves as a nearly-perfect fluid despite the short time scale of the collision. Comparison of experimental data and hydrodynamic simulations \cite{Romatschke:2007mq} indicates that shear viscosity, which takes account of the off-equilibrium corrections from the strong-coupling limit, is extremely small and comparable to the universal lower bound conjectured by the gauge-string correspondence picture \cite{Kovtun:2004de}.

\begin{figure}[tb]
\centerline{\includegraphics[width=12cm,bb=0 0 1660 408]{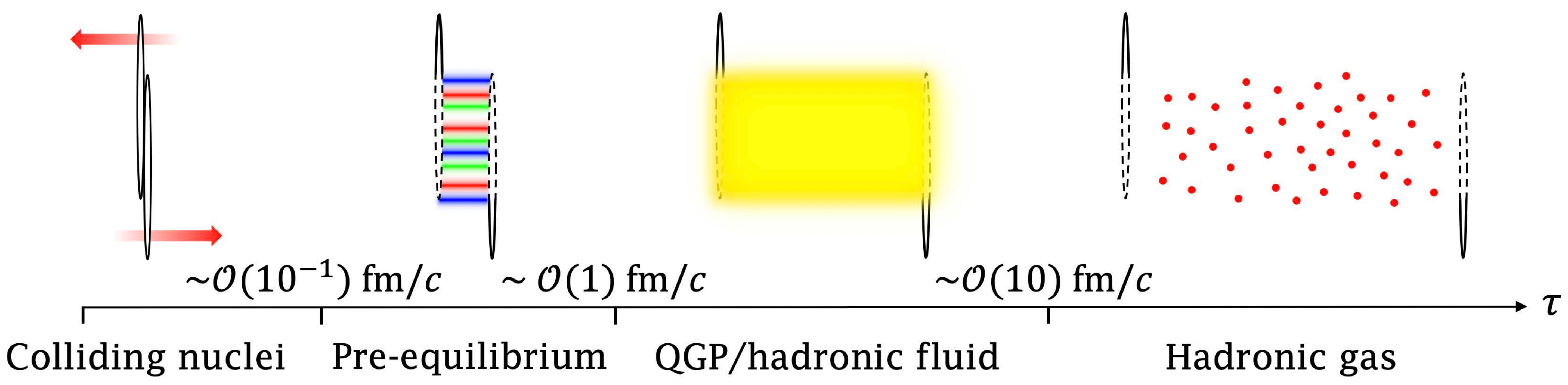}}
\caption{A schematic illustration of the time evolution in the hybrid modeling of relativistic heavy-ion collisions.\label{fig:hic_time}}
\end{figure}

The standard model of relativistic heavy-ion collisions consists of several stages (Fig.~\ref{fig:hic_time}). Two colliding nuclei are quantified as the color glass condensate \cite{McLerran:1993ni,McLerran:1993ka,Iancu:2003xm,Gelis:2010nm,Kovchegov:2012mbw}, a state of gluon saturation, when the collision energy is above around 100 GeV/A. The system after the collision, referred to as glasma, takes early thermalization or \textit{hydrodynamization}, whose precise mechanism is still not well known. Here, hydrodynamization is a concept that the system behaves as a fluid independent of thermalization, which is supported by pressure anisotropy analyses in holographic model and effective kinetic theory. The color glass condensate model implies that color flux tubes are formed in this stage. The turbulent thermalization picture indicates that parton distributions are described by a universal self-similar scaling law before full thermalization. Successful application of the hydrodynamic model to the nuclear collisions requires the pre-equilibrium stage to end before 1 fm/$c$, imposing a tight constraint on the local hydrodynamization time. The typical time scale of the hydrodynamic stage is $\mathcal{O}(1)$-$\mathcal{O}(10)$ fm/$c$. The quark-gluon plasma transit across the crossover temperature $T_c$ into hadrons in this stage as the system expands. The hadronic fluid is then converted into the hadronic gas at kinetic freeze-out, which is sometimes called particlization when the hadronic transport model is employed immediately after the hydrodynamic model. The system is no longer thermal and unstable hadrons decay into stable ones in the post-hydrodynamic stage. 

Photons and leptons are important observables in relativistic heavy-ion collisions \cite{Shuryak:1978ij,Kajantie:1981wg,McLerran:1984ay}. They participate in the collision process only through the electromagnetic and weak interactions as they are colorless particles. The transparency of the QCD matter against those particles makes them work as probes in which the information of the space-time evolution is embedded. Photons and dileptons are thus referred to as electromagnetic probes. 

The photons measured in the experiments, or \textit{inclusive photons}, consist of \textit{direct photons} and \textit{decay photons}. The former is directly produced from the QCD matter during its space-time evolution while the latter is produced as decay products of hadrons such as $\pi^0$ and $\eta$ after the system is decoupled. Decay photon spectra thus reflect the hadronic spectra. In order to probe the evolving QCD matter, direct photons have to be separated from decay photons. It has not been until recently that direct photons and their momentum anisotropies have become measurable with accuracy in high-energy heavy-ion collisions.

\begin{figure}[tb]
\centerline{\includegraphics[width=11cm,bb=0 0 1742 864]{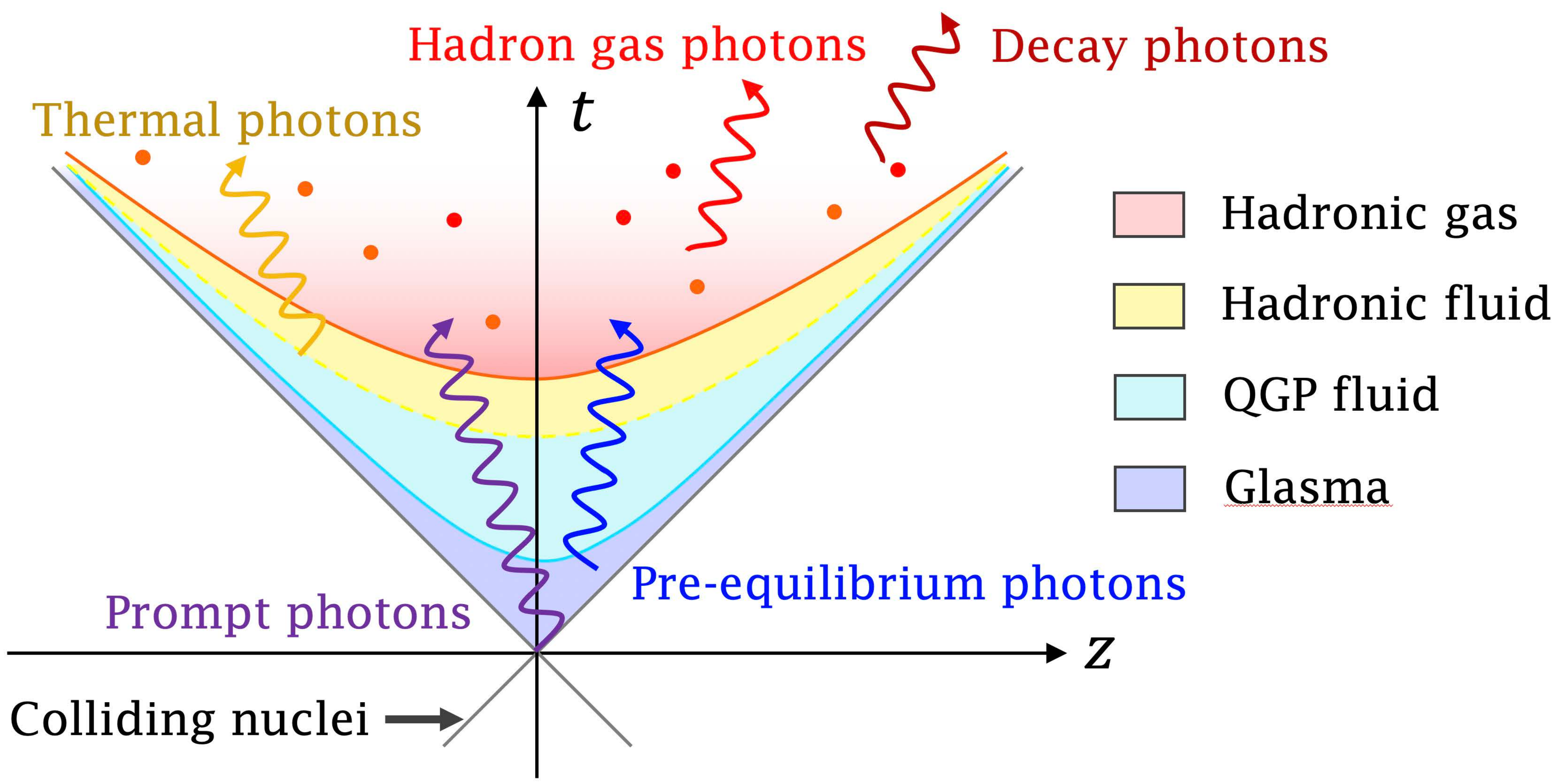}}
\caption{A schematic illustration of the photon emission from different stages in a relativistic heavy-ion collision.\label{fig:hic_photons}}
\end{figure}

There are several sources of direct photons corresponding to different stages in the model of heavy-ion collisions (Fig.~\ref{fig:hic_photons}). Prompt photons, also referred to as perturbative QCD (pQCD) photons, are the most hard contributions produced at the time of the collision. Pre-equilibrium photons are supposed to be semi-hard photons emitted from the glasma. The details are still not well-known owing to the ambiguities in the description of thermalization/hydrodynamization. Thermal photons are soft contributions produced by the QGP and hadronic fluid. Collective dynamics is primarily encoded in the thermal photons. In addition, hadronic gas photons could be emitted during the hadronic transport processes. The difference between those photons and decay photons is the time scale; the former are produced while the system is still interacting, albeit weakly, and the latter includes the photons produced by the decay of hadronic particles after they are decoupled. There can also be other sources of photons which may well be important; notable examples include the photons originated from mini-jets and initial magnetic fields. 

One of the most successful experimental achievements of direct photon analyses would be the extraction of the effective temperature of the hot QCD matter from the transverse momentum spectra \cite{PHENIX:2008uif}. Functional fitting analyses imply that the effective temperatures at RHIC \cite{PHENIX:2008uif,PHENIX:2014nkk} and LHC \cite{ALICE:2015xmh} are both well above the crossover temperature of $T_c \sim 155$-$160$ MeV estimated in (2+1)-flavor lattice QCD simulations \cite{Aoki:2009sc, Bazavov:2011nk, Borsanyi:2013bia, Bazavov:2014pvz}, which is considered as a strong evidence that the deconfined quark matter is created in the heavy-ion collisions. It should be noted that the effective temperatures are averaged over volume and time. The local temperature can be larger at initial times because of medium expansion and at hot spots because of fluctuating initial geometries. 

The flow harmonics of direct photons have attracted much attention since the observed elliptic flow has been discovered to be larger than those predicted by most hydrodynamic models \cite{PHENIX:2011oxq,PHENIX:2015igl,ALICE:2018dti}. Triangular flow of direct photons is later found to follow the trend. This problem is known as the \textit{photon puzzle} and has been actively investigated to achieve simultaneous description of the direct photon and hadronic production in heavy-ion collisions.

In this review, we will discuss direct photon estimations in the hydrodynamic modeling of heavy-ion collisions after a status summary of the theoretical and experimental studies in the heavy-ion community. We consider the interplay of prompt, pre-equilibrium, and thermal photons in relativistic nuclear collisions and its effect on the transverse momentum spectra and elliptic flow coefficients for demonstrative purposes \cite{Monnai:2019vup,Monnai:2018eoh}. Prompt and thermal photons have conventionally been considered as the sources of direct photons, though it has been pointed out that the glasma can also participate in the emission process. We will parametrically scale the bottom-up thermalization process \cite{Baier:2000sb} with turbulent thermalization \cite{Berges:2017eom} into the pre-equilibrium stage before the hydrodynamic evolution to demonstrate if pre-equilibrium contributions can be comparable to the other contributions. 
Reviews on direct photon production can also be found, for example, in Refs.~\citen{Ferbel:1984ef,Alam:1996fd,Cassing:1999es,Peitzmann:2001mz,Gale:2003iz,Rapp:2004qs,Stankus:2005eq,Kapusta:2006pm,David:2006sr,Gale:2009gc,Sakaguchi:2014ewa,David:2019wpt}.

The natural units $c=\hbar=k_B=1$ and the time-like convention of the Minkowski metric $g^{\mu \nu} = \mathrm{diag}(+,-,-,-)$ will be used in the review.

\section{Status}

We review the status of theoretical estimations of direct photon production in high-energy nuclear collisions along with experimental developments. Our focus will be on the phenomenology of the momentum-dependent yields of direct photons in the context of relativistic hydrodynamic modeling. The derivation of direct photon emission rates from individual sources will be discussed elsewhere.

\subsection{Transverse momentum spectra}

Direct photons have been an important observable in relativistic heavy-ion collisions because the existence of thermal photons can be a direct evidence of the QGP creation. The observation at the CERN Super Proton Synchrotron (SPS) indicted that both QGP and non-QGP scenarios could explain the WA98 experimental data of central Pb+Pb collisions at $\sqrt{s_{NN}}= 17.3$ GeV \cite{WA98:2000vxl}. A clear signal of direct photon excess, which would be attributed to thermal sources, was found at RHIC in Au+Au collisions at $\sqrt{s_{NN}}= 200$ GeV \cite{PHENIX:2008uif,PHENIX:2014nkk}. The baseline has been estimated by scaling a functional fit to the direct photon spectra of $p$+$p$ collisions, which is known to be mostly consistent with the pQCD description in the measured $p_T$ range. The inverse slope parameter $T$ is identified as an effective temperature  and is extracted by fitting an exponential function $A \exp(p_T/T)$ to the direct photon spectra after subtracting the prompt photon component assuming the rest represents thermal photons. It has been pointed out that inverse slope is independent of centrality classes within uncertainties;  $T = 239\pm 25^\mathrm{stat} \pm 7^\mathrm{syst}$ MeV for 0-20\%, $260\pm 33^\mathrm{stat} \pm 8^\mathrm{syst}$ MeV for 20-40\%, $225\pm 28^\mathrm{stat} \pm 6^\mathrm{syst}$ MeV for 40-60\%, and $238\pm 50^\mathrm{stat} \pm 6^\mathrm{syst}$ MeV for 60-92\% \cite{PHENIX:2014nkk}. A similar inverse slope analysis suggests higher effective temperatures of $T = 297\pm 12^\mathrm{stat} \pm 41^\mathrm{syst}$ for 0-20\% centrality events and $410\pm 84^\mathrm{stat} \pm 140^\mathrm{syst}$ for 20-40\% centrality events, with larger uncertainties, at $\sqrt{s_{NN}}= 2.76$ TeV Pb+Pb collisions in LHC \cite{ALICE:2015xmh}.

The inverse slope parameter, on the other hand, is not a direct representative of the medium temperature. Besides the obvious spatial and temporal inhomogeneity of the system, the radial flow effect is known to modify the particle spectra \cite{Shen:2013vja}. It should be noted that the blue-shifting effects of hadrons and photons are not the same, because the photons from the away-side fluid elements are red-shifted and partially cancels the blue-shifting effect owing to the transparency of the QCD medium against photon propagation. The thermal photon emission rate is also not exponential -- it would have an additional logarithmic factor $\log(p_T/T)$ in the QGP phase when the reactions $q\bar{q} \to g\gamma$, $qg \to q\gamma$, and $\bar q g \to \bar q \gamma$ are considered \cite{Kapusta:1991qp,Baier:1991em} and a more complicated dependence on $p_T$ and $T$ in the hadronic phase \cite{Turbide:2003si}. Thus, hydrodynamic analyses are required for quantitative extraction of the medium properties via direct photons.

Direct photons have been investigated quantitatively since the hydrodynamic simulation of relativistic nuclear collisions became available. Thermal photon contributions are usually estimated by the integration of the thermal emission rate over the space-time volume using the flow profile for the estimation of Lorentz boost and combined with non-thermal photon contributions such as prompt photons. The transverse momentum spectra have been calculated, for instance, in Refs.~\citen{Alam:2000bu,Srivastava:2000pv,Huovinen:2001wx,Turbide:2003si,dEnterria:2005jvq,Liu:2008eh} using the ideal hydrodynamic model. The initial temperature is estimated to be 300-600~MeV, where the temperature range comes in part from the choice of the initial time of hydrodynamic evolution.

\subsection{Flow harmonics}

Elliptic flow of direct photons is an important observable because the development of azimuthal momentum anisotropy, and thus possibly the origin of nearly perfect fluidity, can be encoded in the observable. Thermal photons are the primary source of momentum anisotropy of direct photons in the conventional picture. The hydrodynamic model has been used to make theoretical estimations of direct photon elliptic flow, which have been improved with the advent of viscous and event-by-event hydrodynamic simulations in early 2010's.

\begin{figure}[tb]
\centerline{\includegraphics[width=3.0in,bb=0 0 567 497]{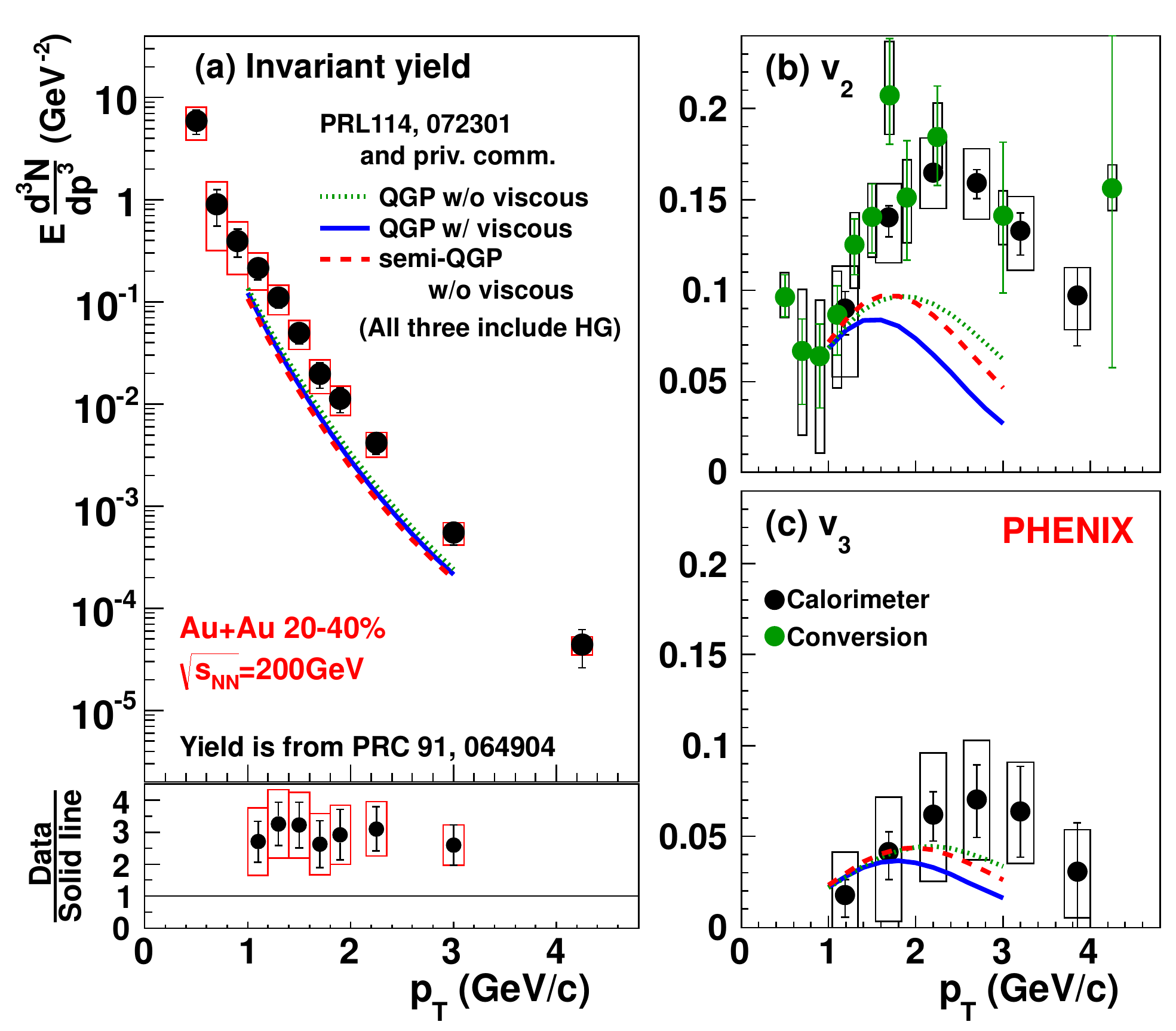}}
\caption{(a) $p_T$ spectra, (b) elliptic flow $v_2$, and (c) triangular flow $v_3$ of direct photons by PHENIX Collaboration at RHIC. The figures are from Ref.~\citen{PHENIX:2015igl}. Theoretical estimations are originally from Refs.~\citen{Gale:2014dfa,Paquet:2015lta}. \label{fig:PHENIX_exp}}
\end{figure}

\begin{figure}[tb]
\centerline{\includegraphics[width=2.2in,bb=0 0 567 650]{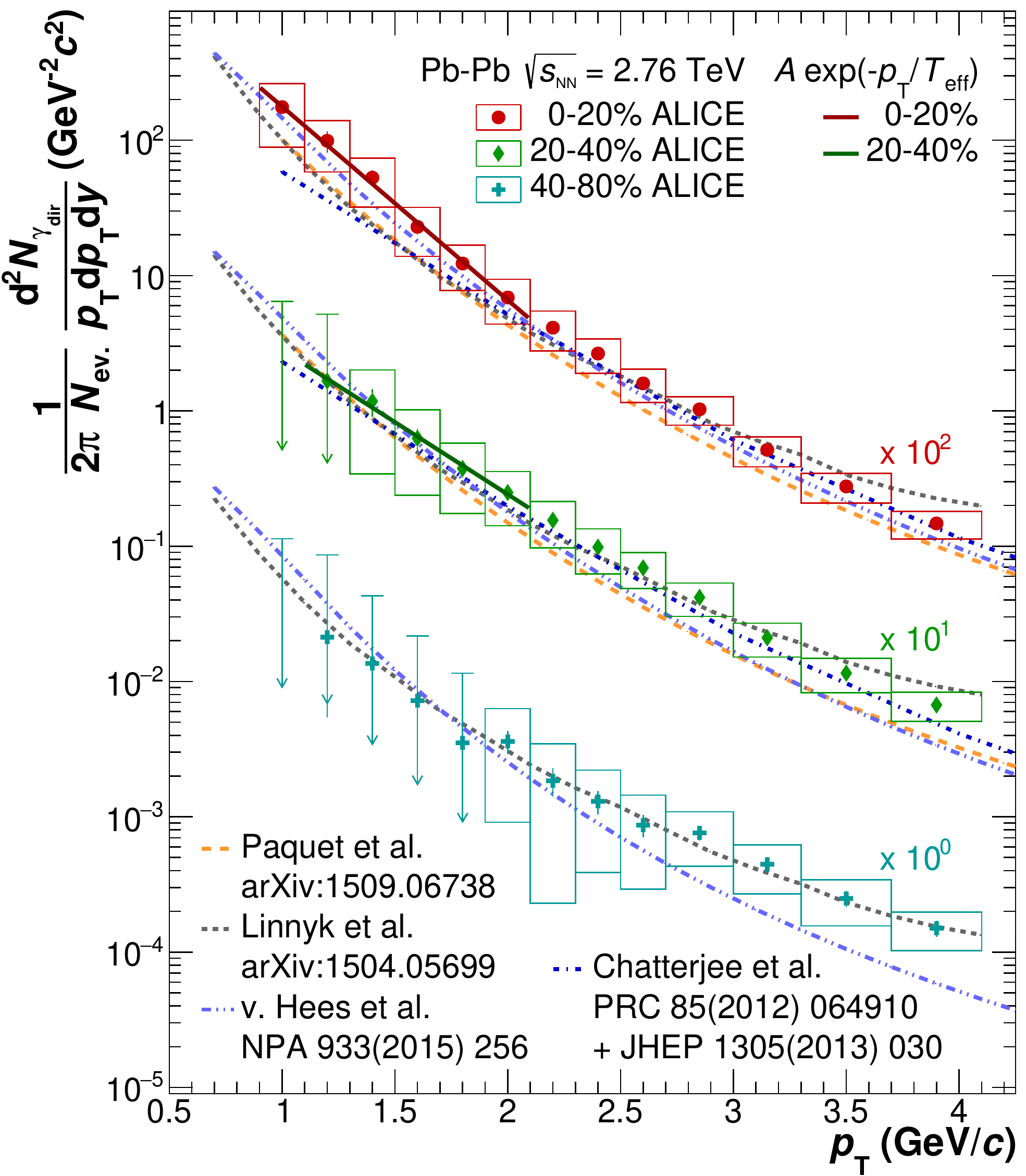} \ \
\includegraphics[width=2.6in,bb=0 0 567 551]{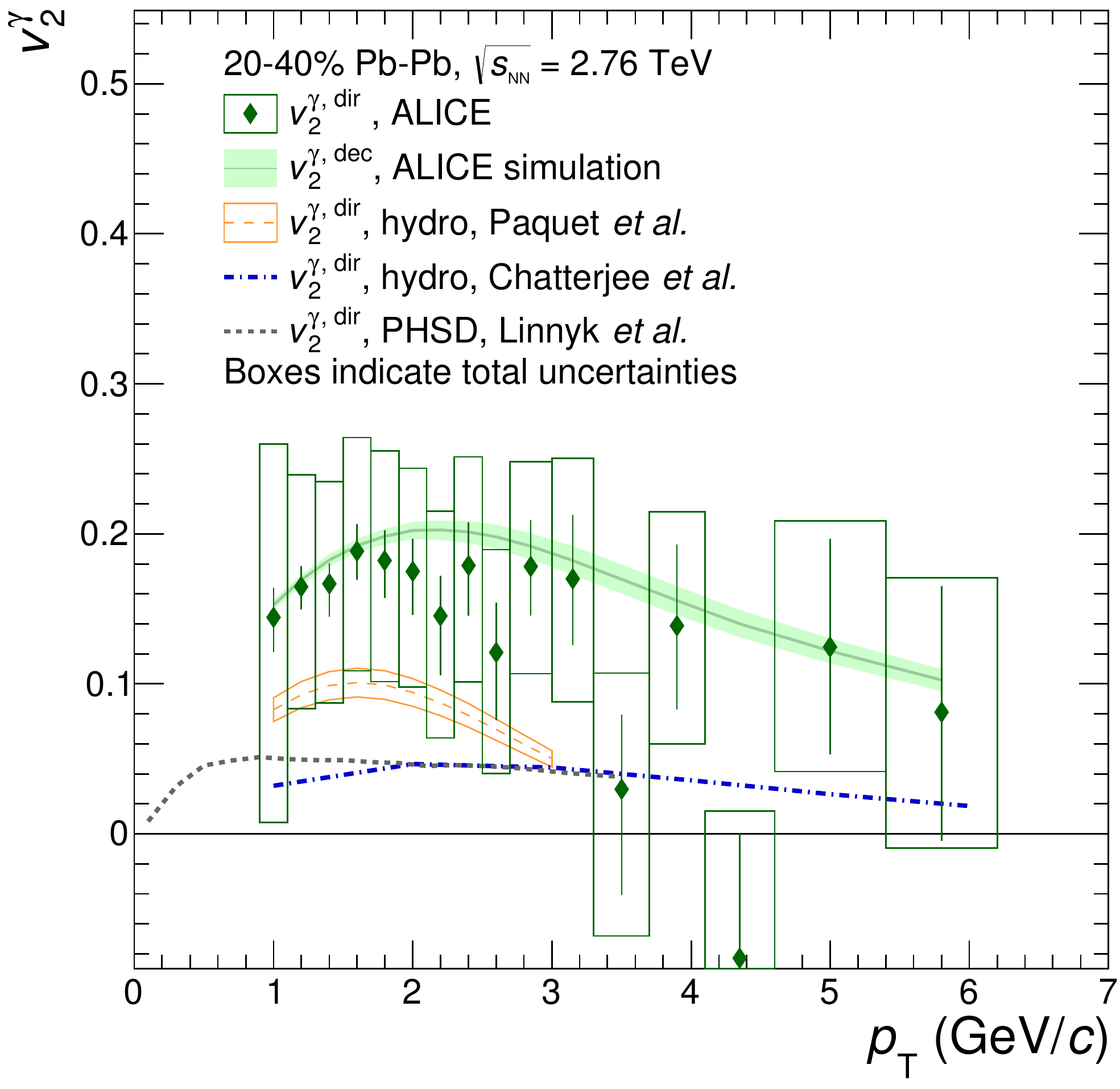}}
\caption{(Left) Direct photon $p_T$ spectra and (Right) elliptic flow $v_2$ by ALICE Collaboration at LHC. The figures are from Refs.~\citen{ALICE:2015xmh,ALICE:2018dti}. Theoretical estimations are originally from Refs.~\citen{Gale:2014dfa,Linnyk:2015tha,Chatterjee:2017akg}. \label{fig:ALICE_exp}}
\end{figure}

The measurement of direct photon $v_2$, on the other hand, is known to be an experimentally difficult task in part because of the large decay photon contributions. An early attempt to extract the quantity from inclusive photon and neutral pion measurements seemed to suggest a small $v_2$ \cite{PHENIX:2005xor}.
More precise measurements in $\sqrt{s_{NN}} = 200$ GeV Au+Au collisions by PHENIX Collaboration at RHIC (Fig.~\ref{fig:PHENIX_exp}) has revealed differential elliptic flow of direct photons to be as large as those of neutral pions \cite{PHENIX:2011oxq,PHENIX:2015igl}, which has been a surprise because unlike hadrons, photons have contributions from early stages in hydrodynamic evolution where the azimuthal momentum anisotropy has not developed yet. Most hydrodynamic simulations thus systematically undershoot the experimental data roughly by a factor of two around $p_T \sim$ 2-3 GeV and much efforts have been made to improve the situation \cite{Chatterjee:2005de, Chatterjee:2008tp, Holopainen:2011pd, vanHees:2011vb, Shen:2013cca, Shen:2013vja, Paquet:2015lta, Kim:2016ylr, Chatterjee:2017akg}. Direct photon elliptic flow observed by ALICE Collaboration at LHC in $\sqrt{s_{NN}} = 2.76$ TeV Pb+Pb collisions also exhibits a similar trend, though it is not completely clear possibly owing to uncertainties (Fig.~\ref{fig:ALICE_exp}) \cite{ALICE:2018dti}. The situation has been recognized as the \textit{photon puzzle}. 

Relativistic hydrodynamic simulations tend to systematically undershoot the $p_T$ spectra by a factor, albeit within uncertainties, which is often considered to be a part of the puzzle. It has also been found that direct photon triangular flow $v_3$ is large and comparable to the pionic counterpart \cite{Mizuno:2014via,PHENIX:2015igl}, which implies that the large anisotropies of direct photons originate from the properties of the medium itself.  

\subsection{Photon puzzle}

\begin{figure}[tb]
\centerline{\includegraphics[width=2.4in,bb=0 0 567 441]{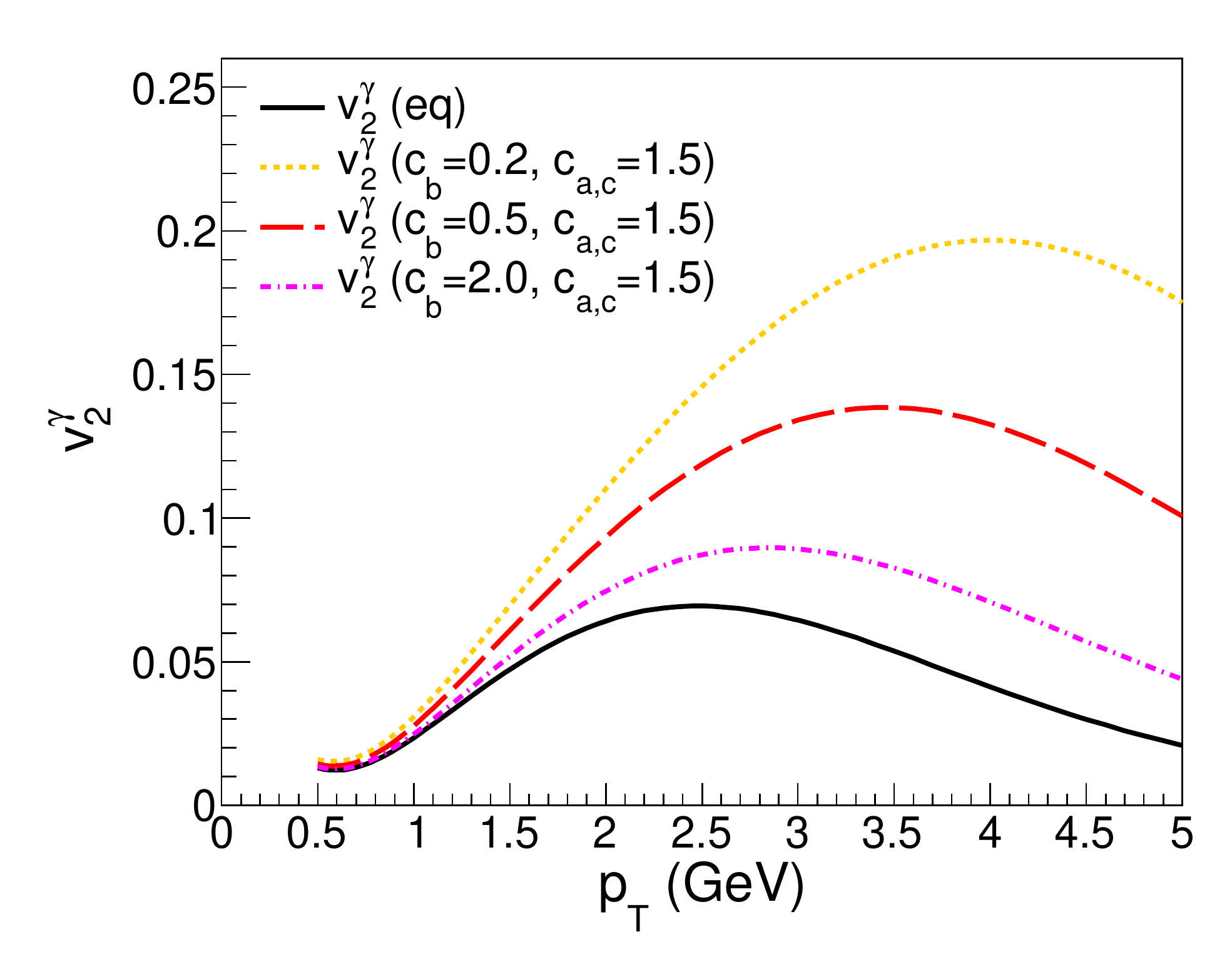}
\includegraphics[width=2.4in,bb=0 0 567 441]{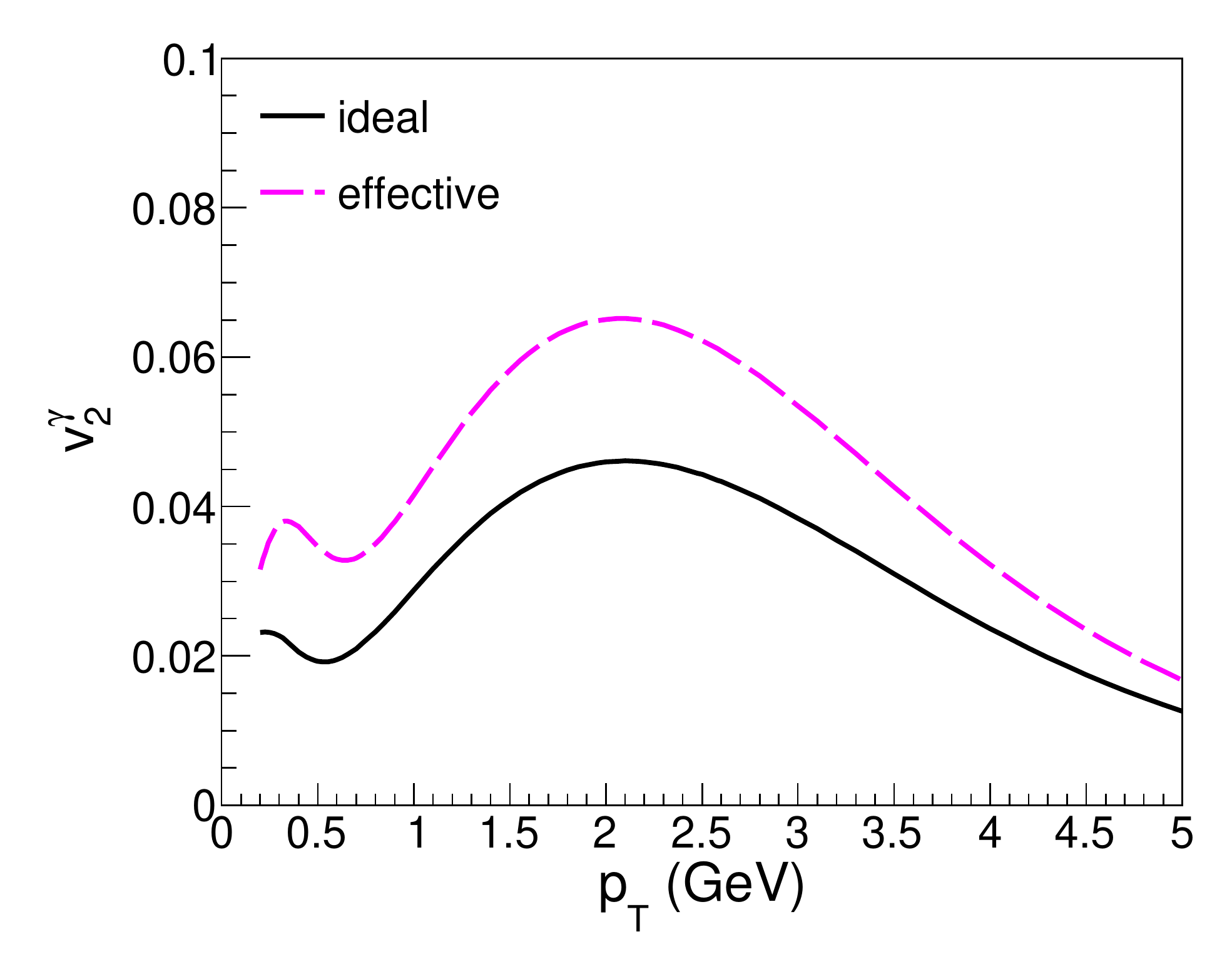}}
\centerline{\includegraphics[width=2.4in,bb=0 0 567 441]{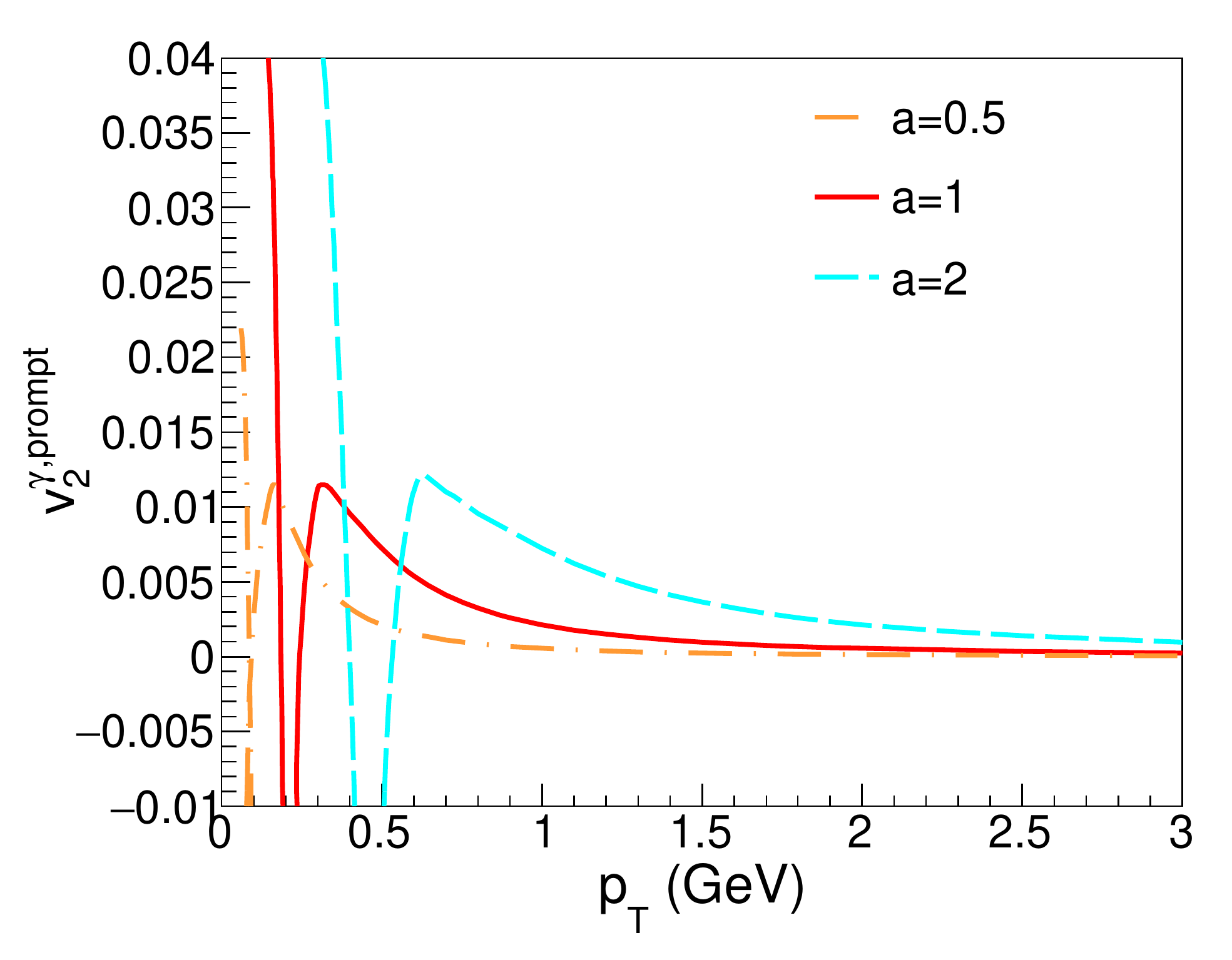}
}
\caption{Examples of theoretical scenarios for producing extra direct photon anisotropy: (top left) thermal photon $v_2$ from a chemically-equilibrating quark matter at different reaction rates \cite{Monnai:2014kqa,Monnai:2015qha}, (top right) thermal photon $v_2$ with in-medium effective phase-space distributions \cite{Monnai:2015bca}, and (bottom) prompt photon $v_2$ induced by lensing effect of the medium with different refraction indices \cite{Monnai:2014taa}.
\label{fig:theory_v2}}
\end{figure}

Various scenarios have been proposed so far to solve the photon puzzle in relativistic nuclear collisions. A few examples can be found in Fig. \ref{fig:theory_v2} (and also in Figs.~\ref{fig:PHENIX_exp} and \ref{fig:ALICE_exp} as theoretical estimations). They are broadly categorized into the modification of the description of bulk medium and the improvement of the photon emission mechanisms. The former includes introduction of shear \cite{Romatschke:2007mq} and bulk \cite{Song:2009rh} viscous corrections, event-by-event initial geometries \cite{Gyulassy:1996br,Osada:2001hw,Aguiar:2001ac}, and initial flow in the relativistic hydrodynamic model. Inclusion of viscosity in the model \cite{Dion:2011pp} requires implementation of viscous correction to the photon emission rate for consistency. This is a non-trivial issue because (i) the transport coefficients have uncertainties, (ii) there are many photon emission processes to introduce the viscous correction, and (iii) the off-equilibrium phase-space distribution used in the derivation is still a topic of debate \cite{Teaney:2003kp,Monnai:2009ad,Denicol:2009am,Monnai:2010qp}. None of the difficulties has been fully solved so far. Also, shear viscosity is known to decrease elliptic flow and bulk viscosity has a relatively small effect except in the vicinity of $T_c$. 
On the other hand, geometrical fluctuation in the initial condition has been reported to enhance direct photon $p_T$ spectra \cite{Chatterjee:2011dw,Chatterjee:2012dn} and elliptic flow \cite{Chatterjee:2013naa}.
The numerical results of a state-of-the-art hydrodynamic model, which provides good description of the hadronic observables, may still be systematically small compared with the data, though the discrepancy is much reduced \cite{Paquet:2015lta}. 

The modification of the direct photon emission mechanism has also been investigated extensively. The large azimuthal momentum anisotropy na\"{i}vely hints that thermal photon emission would be suppressed at early times and/or enhanced at late times. One such possible scenario is the slow quark production in the QGP \cite{Gelis:2004ep,Liu:2012ax,Monnai:2014xya,Monnai:2014kqa,Vovchenko:2016ijt} because the quark-antiquark pair annihilation and (anti)quark-gluon Compton scatterings are the primary sources of thermal photon radiation. The color glass condensate picture implies the existence of a gluon-rich system at initial times, and local chemical equilibration is typically slower than local thermalization because the former is driven only by inelastic processes while the latter by both elastic and inelastic processes. Another scenario is the semi-QGP \cite{Pisarski:2006hz}, where the partially deconfined quark matter is considered in the temperature near and above $T_c$ instead of the perturbative QGP. This leads to the suppression of thermal photon radiation in those temperature range, causing the enhancement of thermal photon elliptic flow \cite{Gale:2014dfa}. More phenomenologically, the deviation of the strongly-interacting QGP from the free parton gas picture may be quantified using the quasi-particle model constrained by the lattice QCD equation of state \cite{Monnai:2015qha}. Those mechanisms tend to enhance direct photon $v_2$ at the cost of the reduction in yields. 

The conventional models consider prompt and thermal photons for estimation. On the other hand, the hybrid model of relativistic heavy-ion collisions implies that pre-equilibrium and post-equilibrium (hadron gas) photons ought to be taken into account for a comprehensive picture. The photon emission in the pre-equilibrium phase has been discussed extensively \cite{Ornik:1995yk,Wang:1996yf,Blaizot:2011xf,Chiu:2012ij,McLerran:2014hza,Greif:2016jeb,Berges:2017eom,Oliva:2017pri,Khachatryan:2018ori,Oliva:2019plz}. 
The implementation of pre-equilibrium photons in the hydrodynamic model has difficulties because, as previously mentioned, early thermalization/hydrodynamization of the strongly-interacting QCD system is not completely understood quantitatively. Early studies, for instance, include the extrapolation of the local temperature backward in time into the pre-equilibrium stage with the Bjorken expansion picture \cite{Shen:2014lye}. The bottom-up thermalization with universal scaling solutions of parton distributions \cite{Berges:2017eom} is parametrically scaled and implemented in the hydrodynamic model in Ref.~\citen{Monnai:2019vup,Monnai:2018eoh}. This will be discussed in Sec.~\ref{sec:3}. A similar approach, but with a cut-off of the pre-equilibrium picture at the hydrodynamization time instead of the uniform scaling, is studied \cite{Garcia-Montero:2019kjk,Garcia-Montero:2019vju} using the hydrodynamic code VISHNU \cite{Shen:2014vra}. 
A solution of the diffusion approximation of the Boltzmann equation is considered \cite{Churchill:2020uvk}
with the hydrodynamic code \textsc{music} \cite{Schenke:2010nt,Schenke:2010rr,Schenke:2011bn} and later updated \cite{Gale:2021zlc} by the K$\o$MP$\o$ST model \cite{Kurkela:2018wud,Kurkela:2018vqr}. The inclusion of the pre-equilibrium photons with zero anisotropy tend to reduce direct photon $v_2$, but retuning of the parameters could lessen the effect.

Late time direct photon emission extends into the off-equilibrium hadron gas phase after particlization. Since the azimuthal momentum anisotropy would be fully-developed in the system, the photon emission from the hadron gas such as $\pi + \pi \to \pi + \pi + \gamma$ may well be important. The late time photon emission scenario has an advantage of increasing $p_T$ spectra and $v_2$ of direct photons simultaneously. It is sometimes phenomenologically imitated by running the hydrodynamic simulation into the low-temperature region after particlization and calculating thermal photons to estimate an upper limit for the correction to $v_2$. In early studies, direct photon particle spectra is estimated using ideal hydrodynamic model and the Ultra-relativistic Quantum Molecular Dynamics (UrQMD) model \cite{Baeuchle:2009ep}.
Hadron gas photons have also been calculated naturally in the direct photon estimation of the Parton-Hadron String Dynamics (PHSD) approach, an off-shell transport model for the hot QCD matter \cite{Linnyk:2015tha,Linnyk:2015rco}. The Simulating Many Accelerated Strongly-interacting Hadrons (SMASH) model \cite{Weil:2016zrk} has been developed recently and used to estimate the hadron gas photons \cite{Schafer:2019edr}, whose results suggest that the late time emission might indeed be important in the framework of the hybrid model \cite{Schafer:2021slz}. It is also proposed that direct photons can be produced during the recombination processes of quarks into hadrons \cite{Fujii:2017nbv}. Possible pseudo-critical enhancement of the thermal photon emission rate has been investigated as a more phenomenological approach \cite{vanHees:2014ida}. 

Other photon production scenarios are also proposed. Jets, for instance, are prominent sources of direct photon emission \cite{Fries:2002kt,Turbide:2005bz,Turbide:2005fk,Turbide:2007mi,Qin:2009bk,Renk:2013kya,Hattori:2016jix}. Jet-medium interaction can induce photons, which are sometimes referred to as jet bremsstrahlung photons, and jets survived through the medium can also produce photons via fragmentation, which can be called jet fragmentation photons. The effects are expected to be larger at higher energies with more jets, and more precise data at the LHC heavy-ion programs would be helpful for distinguishing the scenarios. Jet-photon conversions can be effectively taken into account in the Boltzmann approach to multiparton scatterings (BAMPS) model \cite{Greif:2016jeb}. Another candidate of photon emission is the initial strong magnetic field induced by the spectators \cite{Tuchin:2012mf,Basar:2012bp,Bzdak:2012fr,Muller:2013ila,Yin:2013kya,Ayala:2017vex,Ayala:2019jey,Wang:2020dsr}, which may contribute to the direct photon elliptic flow. Enhancement of triangular flow, on the other hand, would need a different explanation. 
Other approaches include, for instance, introduction of medium refraction \cite{Monnai:2014taa}, Tsallis statistics \cite{McLerran:2015mda}, deceleration processes of colliding nuclei \cite{Koide:2016kpe}, relaxation time dependences \cite{Vujanovic:2016anq}, holographic photon emission rates \cite{Iatrakis:2016ugz}, and anisotropic hydrodynamics \cite{Kasmaei:2019ofu}. 

\subsection{Beam energy scan and small systems}

The collision energy dependences of the direct photon production in nuclear collisions has become a topic of interest again since the beam energy scan programs started at RHIC to explore the dense regime of the QCD phase diagram and to look for the signals of a critical point \cite{Asakawa:1989bq}. It has been pointed out that the hydrodynamic QCD medium would be created even at the collision energies around $\sqrt{s_{NN}} = \mathcal{O}(10)$ GeV, an energy regime previous explored by CERN SPS. The rediscovery is in part because of the development of theoretical methods for describing non-equilibrium dynamics in the QCD matter, such as dissipative hydrodynamic and hadronic transport processes, to observe the signals of local equilibrium that gives rise to hydrodynamic behavior. Hadronic flow anisotropies at lower energies have been analyzed in numerical hydrodynamic simulations \cite{Shen:2012vn,Karpenko:2015xea,Denicol:2018wdp,Du:2019obx,
Wu:2021ypv,Schafer:2021csj}.

Direct photon $p_T$ spectra have been measured in Au+Au collisions at $\sqrt{s_{NN}} = 39$ and 62.4 GeV \cite{PHENIX:2018for}. The results of the collisions at different energies from the beam energy scan to top LHC energies seem to scale as a function of charged particle multiplicity. If the soft radiation at lower energies are identified as thermal photons, it would be consistent with the recent hydrodynamic analyses of hadronic observables that suggests the creation of the hot QCD medium in a wide collision energy range. Corresponding hydrodynamic model analyses are in progress \cite{Gale:2018vuh} and a theoretical prediction based on the transport model has also been made \cite{Endres:2015egk}. 

The system size dependence is also an important subject in the phenomenology of nuclear collisions. It has been experimentally pointed out that the small QGP droplet could be produced in $p$+Au, $d$+Au, and $^3$He+Au collisions at top RHIC energies \cite{PHENIX:2018lia}, despite the historical understanding that the bulk medium would be created only in heavy-ion collisions. The hydrodynamic model is also found to describe the flow harmonics of the small systems reasonably well. The origin of collectivity, on the other hand, is a topic of debate; it can be produced in the finial state interactions in the hydrodynamic phase as in the standard model of heavy-ion collisions, while the color glass condensate theory indicates the existence of initial state momentum correlations which can influence the observables \cite{Dusling:2015gta,Schlichting:2016sqo,Mace:2018vwq}.

Direct photons in $\sqrt{s_{NN}}=200$ GeV $d$+Au collisions at RHIC have been reported to be consistent with the cold nuclear matter description within uncertainties \cite{PHENIX:2012krx}. Similar results are obtained in $\sqrt{s_{NN}}=5.02$ GeV $p$+Pb collisions at LHC, possibly owing to uncertainties \cite{Peresunko:2018eyd,Schmidt:2018ivl,Blau:2019pna}. On the other hand, the experimental data are also compatible with the hydrodynamic description and the effect of thermal photons may become more prominent in the most central collisions \cite{Shen:2015qba,Shen:2016zpp}.

\section{An integrated approach to direct photon production in nuclear collisions} 
\label{sec:3}

We discuss direct photon production in relativistic nuclear collisions including the pre-equilibrium contribution in addition to the prompt and thermal contributions \cite{Monnai:2019vup,Monnai:2018eoh}. Transverse momentum spectra and elliptic flow will be demonstrated in numerical simulations using a relativistic hydrodynamic model.

\subsection{Photon emission model}

Thermal photons are estimated by integrating the emission rate over the space-time volume using the flow and temperature profiles from hydrodynamic model calculations. The thermal photon emission rate is obtained by smoothly connecting the photon emission rates in the QGP and hadronic phases as

\begin{eqnarray}
E \frac{dR^{\gamma}_\mathrm{th}}{d^3p} = \frac{1}{2} [1- f(T)] E \frac{dR^{\gamma}_\mathrm{had} }{d^3p}  + \frac{1}{2}[1+ f(T)] E \frac{dR^{\gamma}_\mathrm{QGP}}{d^3p} , \label{eq:thermal}
\end{eqnarray}
where
\begin{eqnarray}
f(T) = \tanh \bigg( \frac{T-T_c}{\Delta T} \bigg) .
\end{eqnarray}
$T_c = 170$ MeV and $\Delta T = 0.1 T_c$ are the connecting temperature and width, respectively. 

The QGP photon emission rate is based on the perturbative approach by Arnold, Moore and Yaffe \cite{Arnold:2001ba,Arnold:2001ms} which reads
\begin{eqnarray}
E \frac{dR^{\gamma}_\mathrm{QGP}}{d^3p} = \sum_f e_f^2 \frac{T^2}{\pi^2}\alpha_\mathrm{EM} \alpha_s f_q (p) \bigg[ \ln \bigg( \frac{T}{m_\infty} \bigg)+ C_\mathrm{tot} \bigg( \frac{p}{T} \bigg) \bigg] ,
\end{eqnarray}
where the thermal quark mass is given as $m_\infty^2 = 4\pi \alpha_s T^2 /3$. $e_f$ is the charge of the quark flavor $f$. $C_\mathrm{tot}$ includes the contributions of the two-to-two processes for general $p/T$ and of the Landau-Pomeranchuk-Migdal effect on the bremsstrahlung and inelastic pair annihilation. Its parametrization can be found in Ref.~\citen{Arnold:2001ms}.

The thermal photon emission rate in the hadronic phase is based on the massive Yang-Mills theory \cite{Gomm:1984at,Song:1993ae} for the gas of light hadrons $\pi, K, \rho, K^*,$ and $a_1$ \cite{Turbide:2003si,Heffernan:2014mla,Holt:2015cda}. Their numerical results are parameterized by functional fits in the respective references. 

The rate is subject to a Lorentz boost when photons are emitted from a moving fluid element, which induces anisotropy in thermal photon spectra. The local rest frame is given by the flow $u^\mu$ of the hydrodynamic model. The Lorentz boosted photon emission rate is calculated by replacing $E$ with $p^\mu u_\mu$. The thermal photon spectra is thus given as
\begin{eqnarray}
E \frac{dN^{\gamma}_\mathrm{th}}{d^3p} = \int d^4x \bigg[ E \frac{dR^{\gamma}_\mathrm{th}}{d^3p} (p^\mu u_\mu,T) \bigg] ,
\end{eqnarray}
where the hypervolume within a particlization hypersurface is often considered for the space-time integration. In this study, an extended volume will be considered to imitate the contributions of hadron gas photons.

Pre-equilibrium photons are subject to the details of local equilibration mechanisms. Here we extend the Berges-Reygers-Tanji-Venugopalan model \cite{Berges:2017eom} and consider the bottom-up thermalization scenario \cite{Baier:2000sb} to divide the pre-equilibrium stage into three stages: (a) the early stage $c_0  <  Q_s \tau < c_1 \alpha_s^{-3/2}$, (b) the intermediate stage $c_1 \alpha_s^{-3/2} < Q_s \tau < c_2 \alpha_s^{-5/2}$, and (b) the late stage $c_2 \alpha_s^{-5/2} < Q_s \tau < c_3 \alpha_s^{-13/5}$ by phenomenologically scaling the process into the hydrodynamization time scale. Here $Q_s$ is the saturation momentum. The scaling factors $c_0,c_1,c_2$, and $c_3$ are determined so that $c_0 Q_s^{-1} $ and $c_3 \alpha_s^{-13/5} Q_s^{-1}$ are identified with the initial time $\tau_\mathrm{ini}$ and the final (hydrodynamization) time $\tau_\mathrm{hyd}$ of the pre-equilibrium stage. 

In the early stage (a), it is indicated by classical statistical simulations that the parton distributions are attracted to self-similar solutions \cite{Berges:2013eia,Berges:2014bba,Berges:2015ixa,Tanji:2017suk,Berges:2020fwq}. Here, they are phenomenologically parametrized as
\begin{eqnarray}
f_{i} &=& h(p_T) \times (Q_s \tau)^\alpha f_{i}^{s} ((Q_s \tau)^{\beta} p_T,(Q_s \tau)^{\gamma} p_z), \label{eq:preeq_f} \\
f_{i}^{s} (p_T,p_z) &=& A_{i} p_T^{-1} \exp(-p_z^2/\sigma_{z}^2), \label{eq:preeq_fs}
\end{eqnarray}
where the subscript $i = g,q$ denotes gluons and quarks. $p_T$ and $p_z$ are transverse and longitudinal momenta. The exponents of the scaling solution are $\alpha~=~-2/3$, $\beta~=~0$, and $\gamma~=~1/3$. The distribution is cut off for momenta larger than $Q_s$ by the function
\begin{eqnarray}
h(p_T) = \frac{1 - \tanh[(p_T-Q_s)/\Delta p_T]}{2}, \label{eq:preeq_h}
\end{eqnarray}
to reproduce the break-down of the scaling law observed in numerical simulations. $\Delta p_T = 0.1 Q_s$ is the width of this transition region.

Non-universal coefficients $A_i$ and $\sigma_z$ are related to the normalization and the longitudinal anisotropy, respectively. The former can be determined by the condition that the energy density distribution at the end of the pre-equilibrium evolution, including the stages (b) and (c) discussed later, serves as the initial condition for the hydrodynamic model that quantitatively describes the experimental data. The energy density in the pre-equilibrium system is estimated as
\begin{eqnarray}
e &=& \frac{1}{(2\pi)^3} \int_0^\infty 2 dp_z \int_0^{\infty} 2 \pi p_T \sqrt{p_T^2 + p_z^2} (d_g f_g + d_q f_q) , \label{eq:e}
\end{eqnarray}
where the degeneracies are $d_g = 2_\mathrm{spin}\times(N_c^2-1)$ and $d_q = 2_\mathrm{spin}\times 2_\mathrm{q\bar{q}}\times N_c \times N_f$. The quark density at early times is assumed to be suppressed compared with the gluon density by $\alpha_s$ owing to the gluon-rich environment of the color glass condensate. 

$\sigma_z$ is the typical longitudinal momentum of the system and thus should be smaller than the effective temperature of the local thermal system with the same amount of energy density because the longitudinal pressure is known to be smaller than the transverse one in an expanding glasma. It will be left as a parameter in our numerical analyses.

The photon emission rate in the small angle approximation \cite{Berges:2017eom, Blaizot:2014jna} is
\begin{eqnarray}
E \frac{dR^\gamma_\mathrm{a}}{d^3p} &=& \sum_f e_f^2 \frac{4}{\pi^2} \alpha_\mathrm{EM} \alpha_s f_q(p) \log \bigg(1+\frac{2.919}{g^2}\bigg) \int \frac{d^3p'}{(2\pi)^3} \frac{1}{p'} [f_g(p') + f_q(p')], \label{eq:rate_a}
\end{eqnarray}
where the logarithmic factor is chosen in accordance with the thermal rate for the pair annihilation and Compton scattering  \cite{Kapusta:1991qp,Baier:1991em}. 

At the intermediate stage (b), the gluon occupation number becomes smaller than unity and the Debye screening starts to be controlled by soft gluons. We simply assume the expression (\ref{eq:rate_a}) is valid because the total number of gluons is still controlled by hard gluons. At the late stage (c), soft partons dominate and the thermal tail in the phase-space distribution is expected to be developed. The photon emission rate is estimated by smoothly interpolating the thermal and the non-thermal emission rates as
\begin{eqnarray}
E \frac{dR_\mathrm{c}^\gamma}{d^3p} &=& \frac{\tau - \tau_\mathrm{c}}{\tau_\mathrm{th} - \tau_\mathrm{c}} E \frac{dR^\gamma_\mathrm{th} }{d^3p} + \frac{\tau_\mathrm{th} - \tau}{\tau_\mathrm{th} - \tau_\mathrm{c}} E \frac{dR^\gamma_\mathrm{b}}{d^3p} , \label{eq:rate_c}
 \end{eqnarray}
 where $\tau_\mathrm{c} = c_2 \alpha_s^{-5/2} Q_s^{-1}$ is the initial time of the stage (c). 
 
Prompt photons are estimated by scaling the direct photon spectra of $p$+$p$ collisions observed in experiments assuming that thermal and glasma photons are not produced in such systems. Here we use the parametrization
 \begin{eqnarray}
E \frac{dN^\gamma_\mathrm{pr}}{d^3p} &=& 6745 \frac{\sqrt{s}}{(p_T)^5} \frac{N_\mathrm{coll}}{\sigma_{pp}^\mathrm{in}} \frac{\mathrm{pb}}{\mathrm{GeV}^2},\label{eq:prompt}
\end{eqnarray}
where $N_\mathrm{coll}$ is the number of collisions and $\sigma_{pp}^\mathrm{in}$ is the inelastic nucleon-nucleon cross section \cite{Turbide:2003si}. One can alternatively use pQCD results as a baseline. Photon production from the color glass condensate has been extensively studied \cite{Gelis:2002ki,Jalilian-Marian:2005tod,Benic:2016yqt,Benic:2016uku,Benic:2018hvb}. Possible effects of glasma photon production in $p$+$p$ collisions on heavy-ion collisions is discussed, for instance, in Ref.~\citen{Monnai:2019vup}. 

\subsection{Numerical demonstration}

Direct photon emission in Pb+Pb collisions at $\sqrt{s_{NN}} = 2.76$ TeV is estimated in numerical simulations based on a (2+1)-dimensional inviscid relativistic hydrodynamic model \cite{Monnai:2014kqa}. The equations of motion are the energy-momentum conservation $\partial_\mu T^{\mu \nu} = 0$. Net baryon, strangeness, and electric charge are assumed to be negligible at the energy in consideration.

The initial energy density distribution for the pre-equilibrium stage and the number of nucleon-nucleon collisions are estimated using the Monte-Carlo Glauber model \cite{Miller:2007ri} with $\sigma_{pp}^\mathrm{in} = 65$ mb. The initial condition is then event-averaged at the impact parameter $b=4.6$ fm to imitate 0-20 \% centrality events for simplicity. The average number of collisions is $N_\mathrm{coll} = 1256$. The initial and hydrodynamization times are set to $\tau_\mathrm{ini} = 1/Q_s$ and $\tau_\mathrm{hyd} = 0.6$ fm/$c$, respectively. We treat the saturation momentum as a parameter and explore the momentum range of 2.5 GeV $\leq Q_s \leq 3.5$~GeV in numerical simulations following the estimation based on $Q_s^2 \sim A^{1/3} Q_{0}^2 (x_0/x)^\lambda$ where $Q_0 = 1$~GeV, $\lambda = 0.288$, $x_0 = 3\times 10^{-4}$, and $x = 5 \times 10^{-4}$ \cite{GolecBiernat:1998js}. $\sigma_z$ is explored in the range of 0.05 GeV $\leq \sigma_z \leq 0.4$~GeV so that the longitudinal pressure is kept smaller than the transverse one. The expansion is assumed to be longitudinal and boost-invariant. $\alpha_\mathrm{EM} = 1/137$, $\alpha_s = 0.2$, $N_c = 3$, and $N_f = 3$ are used in the estimation.

The energy density distribution at the end of the finial stage of pre-equilibrium evolution is used as the initial condition for hydrodynamic evolution. The overall normalization is determined by the experimental data. The equation of state is constructed by smoothly matching the results of lattice QCD \cite{Bazavov:2014pvz} and hadron resonance gas model \cite{Tanabashi:2018oca} in the vicinity of crossover \cite{Monnai:2019hkn,Monnai:2021kgu}. The kinetic freeze-out temperature \cite{Cooper:1974mv} is $T_f = 0.14$~GeV. Hadronic observables are calculated by considering the effects of hadronic decay. We, on the other hand, take into account the contribution of thermal photons in the hydrodynamic model down to the temperature of $T = 0.11$~GeV to phenomenologically compensate for the lack of hadron gas photons in the current model \cite{Paquet:2015lta}. 

\begin{figure}[tb]
\centerline{\includegraphics[width=2.4in,bb=0 0 567 441]{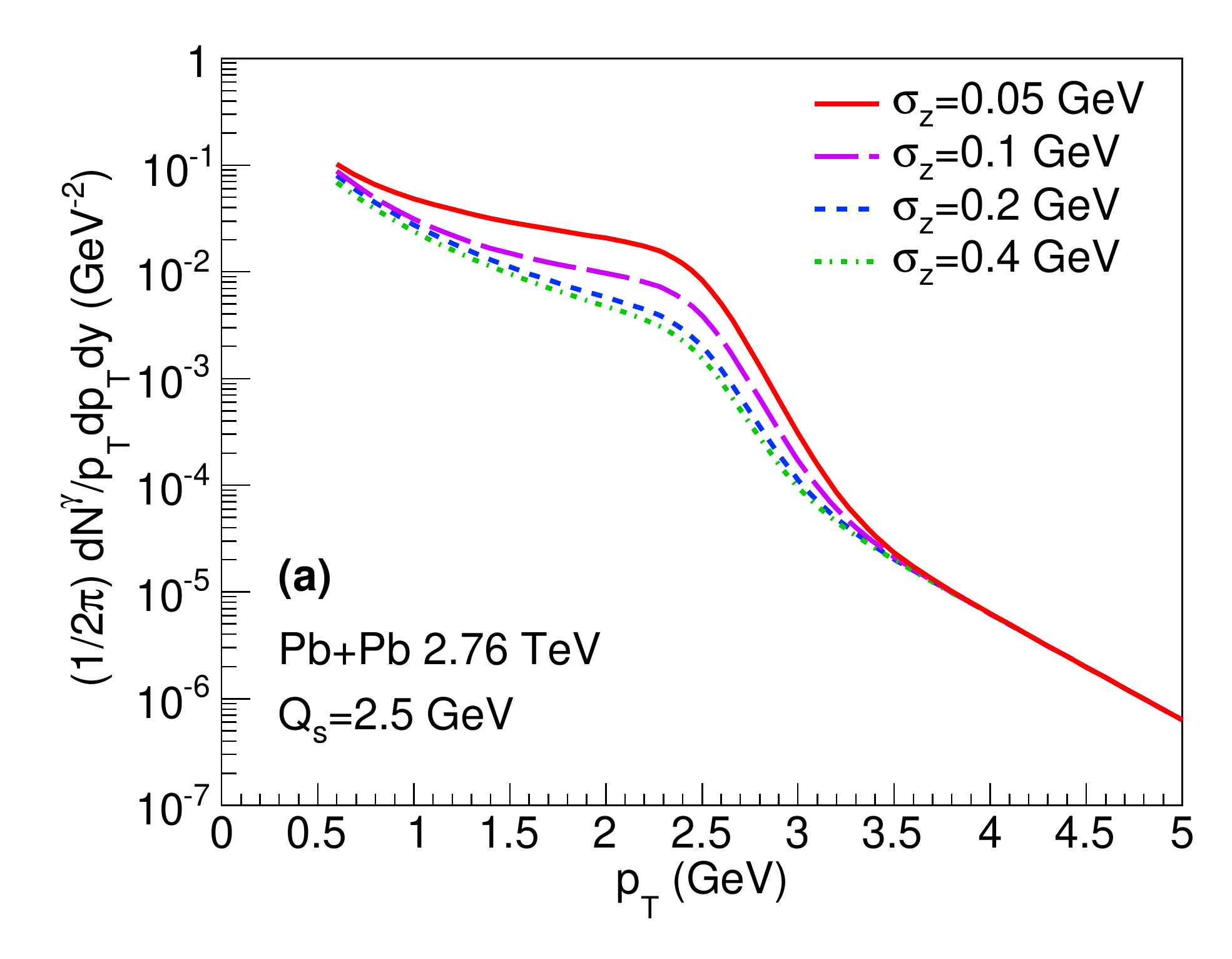}
\includegraphics[width=2.4in,bb=0 0 567 441]{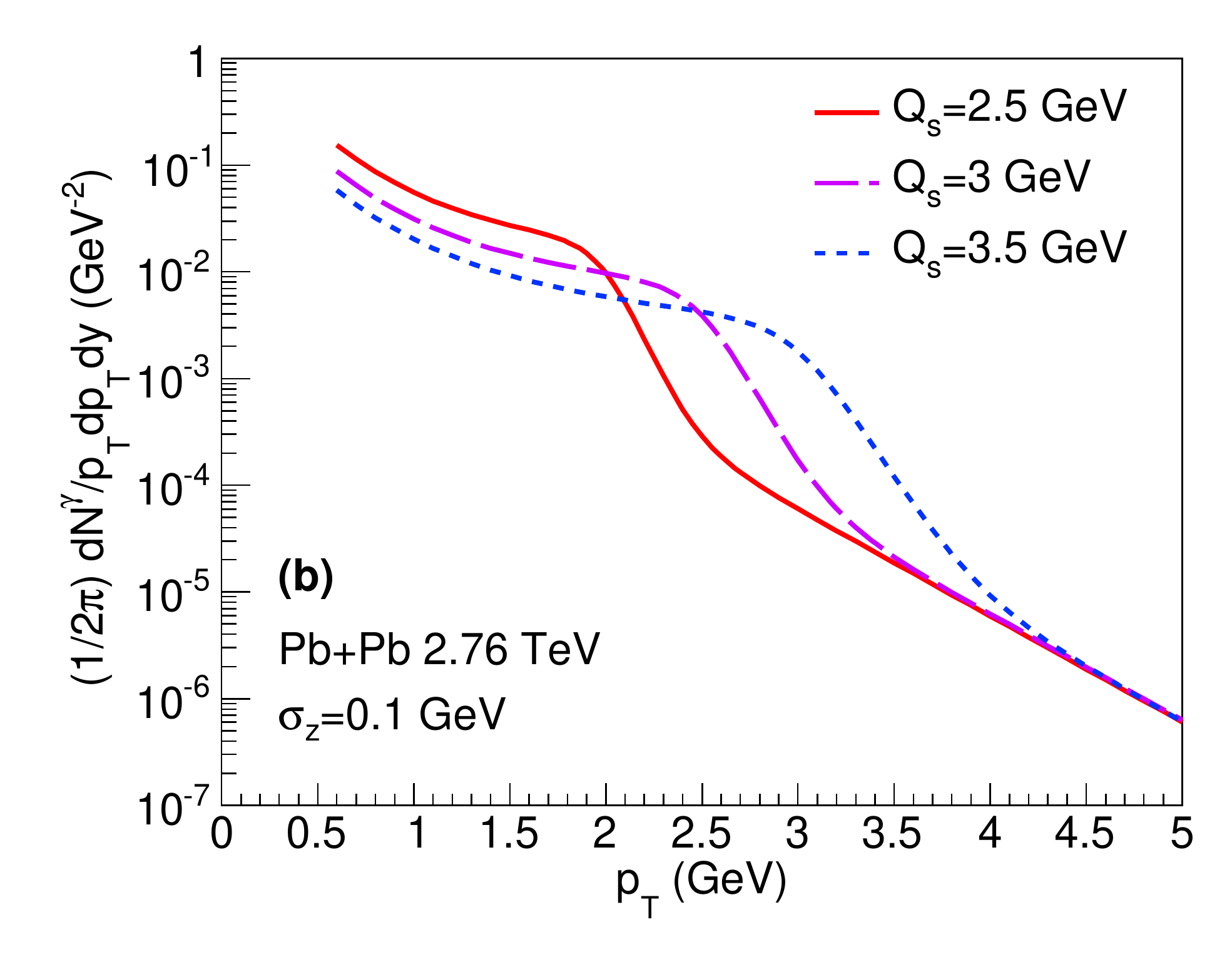}}
\caption{(a) $\sigma_z$ dependence and (b) $Q_s$ dependence of the transverse momentum spectra of pre-equilibrium photons for off-central Pb+Pb collisions at $\sqrt{s_{NN}} = 2.76$ TeV \cite{Monnai:2019vup}.\label{fig:preeq_ph}}
\end{figure}

Transverse momentum spectra of pre-equilibrium photons at midrapidity $y=0$ are shown in Fig.~\ref{fig:preeq_ph}. The distributions have a structure of enhancement around the saturation scale. One can see that smaller $\sigma_z$ leads to larger enhancement  for a fixed $Q_s$ because smaller phase-space volume in the longitudinal direction leads to larger volume in the transverse directions when the local energy density is fixed. Similarly, larger $Q_s$ leads to smaller enhancement at $p_T \sim Q_s$ when $\sigma_z$ is fixed.

\begin{figure}[tb]
\centerline{\includegraphics[width=2.6in,bb=0 0 567 441]{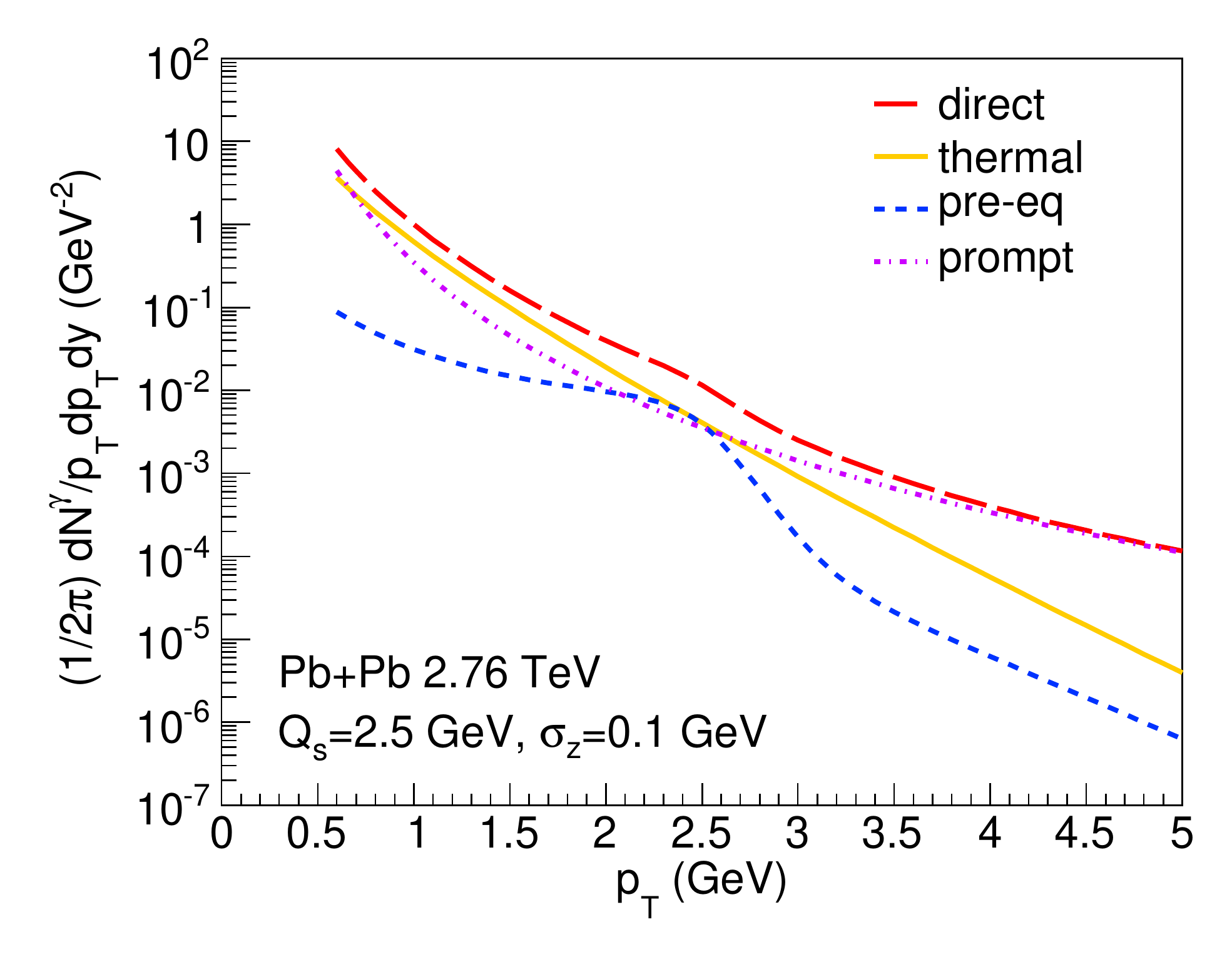}}
\caption{$p_T$ spectra of direct, thermal, pre-equilibrium, and prompt photons at $Q_s = 2.5$ GeV and $\sigma_z = 0.1$ GeV for off-central Pb+Pb collisions at $\sqrt{s_{NN}} = 2.76$ TeV \cite{Monnai:2019vup}.\label{fig:dir_ph}}
\end{figure}

Direct photon $p_T$ spectra at $Q_s = 2.5$ GeV and $\sigma_z = 0.1$ GeV is shown in Fig.~\ref{fig:dir_ph} where prompt, pre-equilibrium, and thermal photon contributions are considered. One can see that pre-equilibrium photons can be non-negligible in comparison to the other types of photons for the chosen set of saturation scale and momentum anisotropy. Prompt photons dominate the mid-high $p_T$ region above $Q_s$ because they are produced in the initial hard processes. Thermal photons, on the other hand, become important in the low $p_T$ region around 1-2 GeV. Pre-equilibrium photons are comparable to the thermal and prompt photons in the intermediate region. This is phenomenologically consistent with the expectations that typical momentum scale of the strongly-interacting system decrease along with the space-time evolution. 

\begin{figure}[tb]
\centerline{\includegraphics[width=2.6in,bb=0 0 567 441]{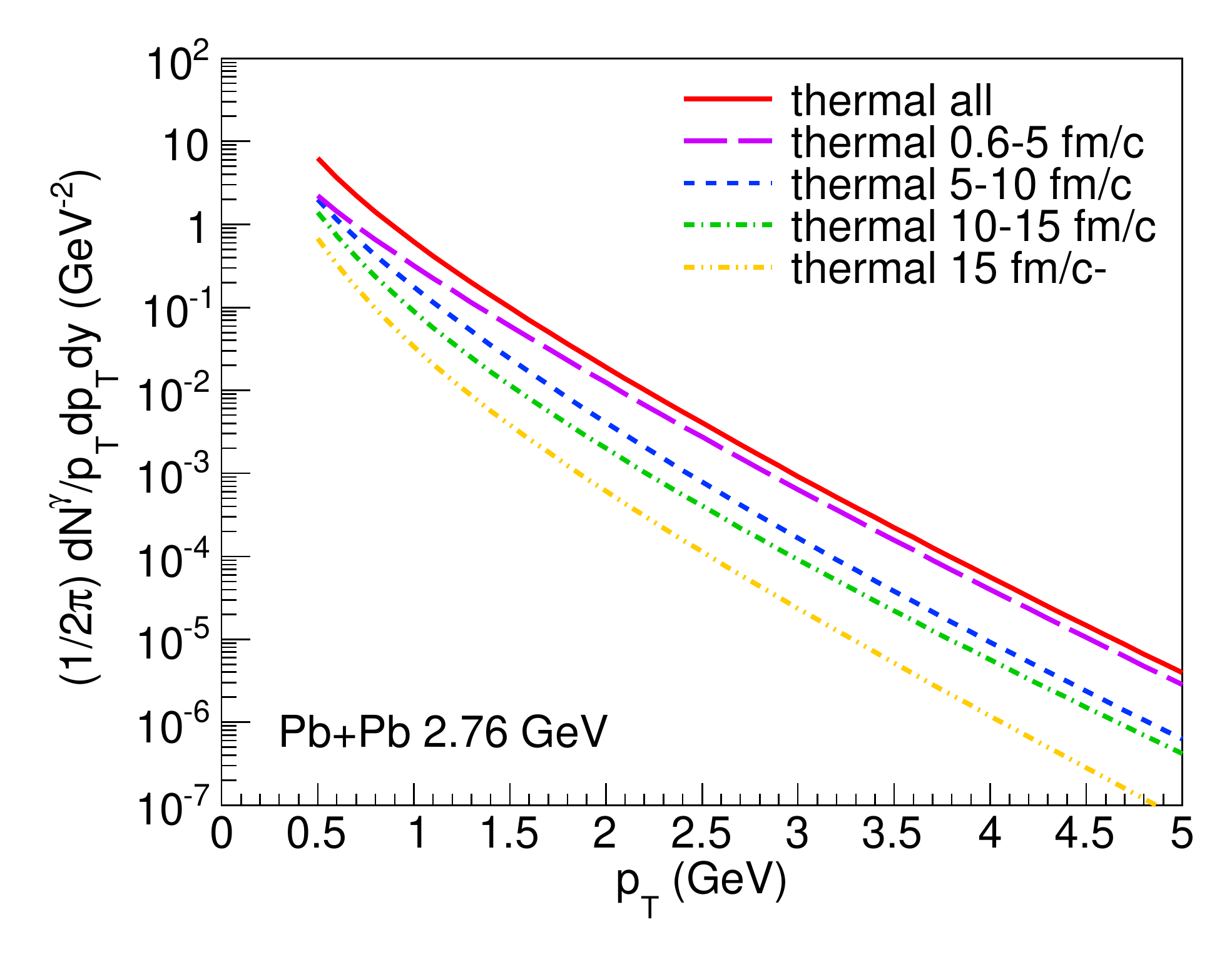}}
\caption{$p_T$ spectrum of thermal photons and its temporal decompositions for off-central Pb+Pb collisions at $\sqrt{s_{NN}} = 2.76$ TeV \cite{Monnai:2019vup}.\label{fig:th_ph}}
\end{figure}

A similar trend can be found when the thermal photon spectra is decomposed into the temporal bins of 0.6-5 fm/$c$, 5-10 fm/$c$, 10-15 fm/$c$, and over 15 fm/$c$ (Fig.~\ref{fig:th_ph}). The photons produced in the late stages of hydrodynamic evolution tend to be soft and have a steeper slope compared with those emitted in the early stages, reflecting the decrease in the average temperature of the medium.

We finally consider the differential elliptic flow of direct photons. Here it is defined simply as the second harmonics of Fourier decomposition 
\begin{eqnarray}
v_2 (p_T,y) &=& \frac{\int d\phi_p \cos[2(\phi_p-\Psi)] \frac{dN^\gamma}{d\phi_p p_T d p_T dy}}{\int d\phi_p  \frac{dN^\gamma}{d\phi_p p_T d p_T dy}} ,
\end{eqnarray}
using the azimuthal momentum angle $\phi_p$ and the event plane angle $\Psi$. 

\begin{figure}[tb]
\centerline{\includegraphics[width=2.6in,bb=0 0 567 441]{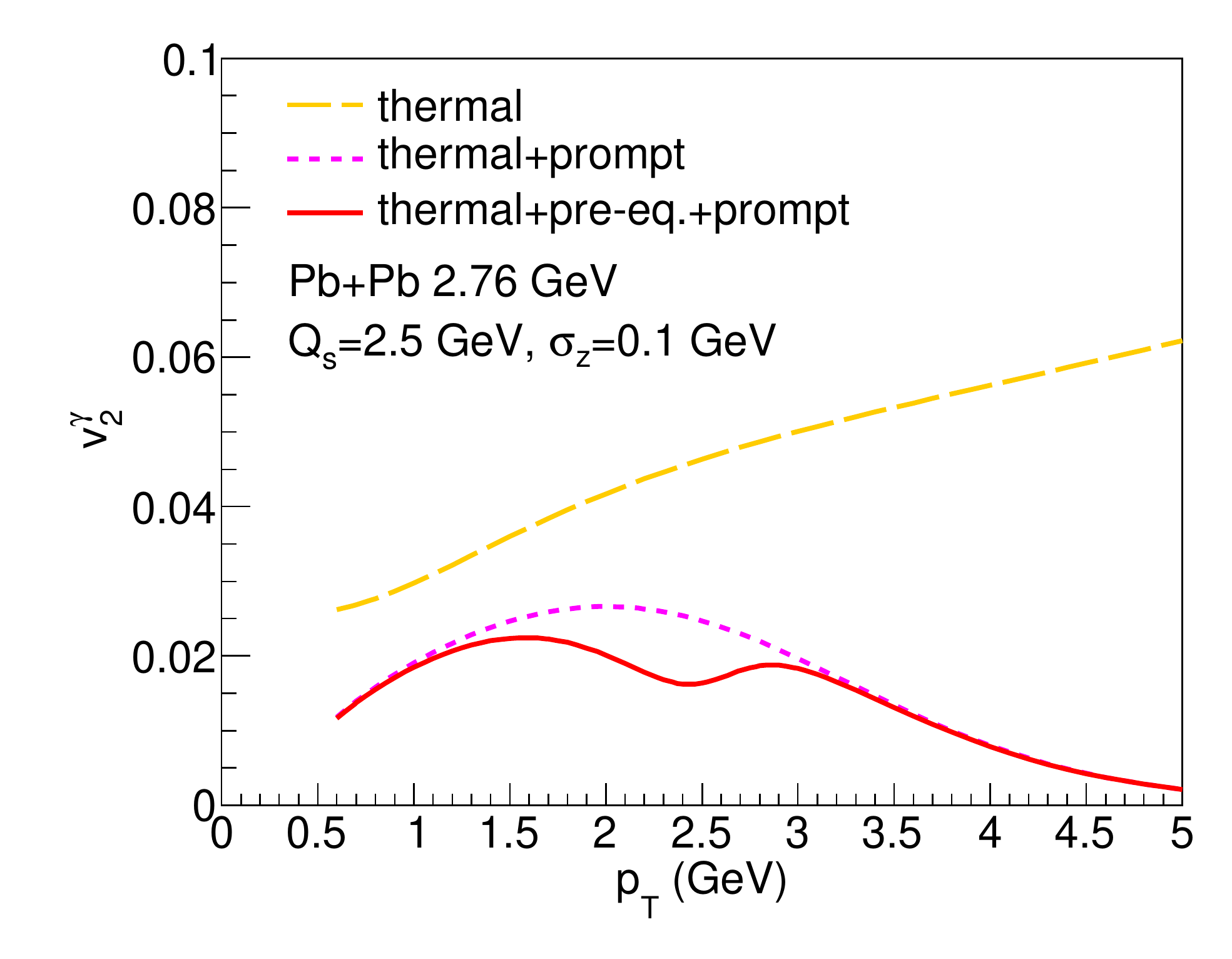}}
\caption{Elliptic flow of thermal photons, thermal and prompt photons, and direct photons for off-central Pb+Pb collisions at $\sqrt{s_{NN}} = 2.76$ TeV \cite{Monnai:2019vup}.\label{fig:dir_v2}}
\end{figure}

The numerical results of direct photon elliptic flow at mid-rapidity are presented in Fig.~\ref{fig:dir_v2}. Thermal photons are diluted by prompt photons with vanishing anisotropy. Pre-equilibrium photons inevitably further decrease direct photon $v_2$ around the saturation momentum scale, assuming that those photons do not have anisotropy. Position-dependent saturation momentum may smear the effect. Also, modification of the hydrodynamization time might minimize the reduction of the flow harmonics.

\subsection{Parameterization of pre-equilibrium photon emission rate}

The pre-equilibrium photon emission rate (\ref{eq:rate_a}) is parameterized here to facilitate implementation in different hydrodynamic models for direct photon estimation. It can be expressed as 
\begin{eqnarray}
E \frac{dR^\gamma_\mathrm{a}}{d^3p} &=& \sum_f e_f^2 \frac{4}{\pi^2} \alpha_\mathrm{EM} \alpha_s f_q(p) \log \bigg(1+\frac{2.919}{g^2}\bigg) (I_g + I_q), \nonumber
\end{eqnarray}
where $f_q$ is defined by Eqs.~(\ref{eq:preeq_f})-(\ref{eq:preeq_h}) and $I_i \ (i=g,q)$ by
\begin{eqnarray}
I_i &=& \int \frac{d^3p}{(2\pi)^3} \frac{1}{p} f_i(p).
\end{eqnarray}
The integrals can be carried out in numerical calculations. Alternatively, they can also be analytically approximated as
\begin{eqnarray}
I_i(\tau, Q_s, \sigma_z) &=& \frac{A_i \sigma_z [(2/3) \log (64 Q_s^4 \tau/\sigma_z^3)+\gamma]}{8\pi^{3/2}Q_s \tau} , 
\end{eqnarray}
for $Q_s > 2$ GeV, $\sigma_z < 0.5$ GeV and $\tau > 0.1$ fm/$c$ where $\gamma$ is Euler's constant. The normalization $A_i$ is constrained by identifying the local energy density of a initial condition model with that calculated using Eq.~(\ref{eq:e}),  which can be approximated as 
\begin{eqnarray}
e (\tau)&=& \frac{(d_g A_g +d_q A_q ) Q_s \sigma_z}{8\pi^{3/2}\tau},
\end{eqnarray}
where $A_q = \alpha_s A_g$. The emission rate in the stage (b) is assumed to have the same functional form as that in the stage (a). The emission rate in the stage (c) is given by the interpolation in Eq.~(\ref{eq:rate_c}).

\section{Summary and outlook}	

We have reviewed the direct photon production in the framework of relativistic hydrodynamic modeling of high-energy nuclear collisions. Direct photons are a useful tool in elucidating the space-time evolution of the QCD matter because the colorless particles are not scattered by the bulk medium after production. The flow harmonics and particle spectra observed at RHIC in $\sqrt{s_{NN}}=200$ GeV Au+Au collisions, on the other hand, have been found to overshoot most theoretical predictions. Those at LHC in $\sqrt{s_{NN}}=2.76$ TeV Pb+Pb collisions behave similarly but the anomaly has not been confirmed owing to large uncertainties. This is known as the photon puzzle and a variety of scenarios have been proposed to solve it.

Direct photons are conventionally calculated assuming that they consist of prompt and thermal photons. On the other hand, the hybrid model of nuclear collisions implies the existence of pre-equilibrium photons as well as hadron gas photons. A comprehensive understanding of the direct photon production would thus require the estimation of those contributions.

We have discussed the estimation of pre-equilibrium photons in the framework of a hydrodynamic model by phenomenologically scaling the bottom-up thermalization process to the time scale of hydrodynamization constrained by the experimental data. Following the turbulent thermalization approach, self-similar parton distribution functions have been parametrically implemented in the calculation of pre-equilibrium photon radiation. First, we have considered the transverse momentum spectra of pre-equilibrium photons as functions of the saturation scale $Q_s$ and the anisotropy parameter $\sigma_z$ in numerical simulations. The spectrum is found to be enhanced around $Q_s$ and may well be comparable to the thermal and prompt photon contributions depending on the value of $\sigma_z$. This suggests that the effects of pre-equilibrium photons might be visible in direct photon spectra and should not be neglected in general. The modification of the slope parameter can also affect the extraction of the effective temperature. Though hadron gas photons have not been fully implemented in the model, their contributions have been partially imitated by taking into account the thermal photon emission outside the freeze-out hypersurface using the hydrodynamic model. Next, we have calculated elliptic flow of direct photons. It is implied that unless additional parameter tuning is performed, the quantity is suppressed around $Q_s$ in the presence of pre-equilibrium photons assuming that they have no anisotropy. 

It would be pivotal for theoretical models to give a quantitative description to direct photon observables because it would open up the possibilities to extract information about the history of the QCD system after little bangs from experimental data. An important step would be to integrate some of the scenarios proposed so far to learn if a better and consistent picture is obtained. Direct photon data with reduced uncertainties over a wide range of collision energies and system sizes would also be useful in distinguishing theoretical scenarios.

\section*{Acknowledgments}

A.M. acknowledges fruitful discussion in Hard Probes 2018 (Aix-les-Bains, France, October 1-5, 2018). The work of A.M. was supported by JSPS KAKENHI Grant Number JP19K14722.

\bibliography{photons}

\end{document}